\documentclass[referee]{mn2e}
\usepackage{graphics}

\def\ltsima{$\; \buildrel < \over \sim \;$}
\def\lsim{\lower.5ex\hbox{\ltsima}}
\def\gtsima{$\; \buildrel > \over \sim \;$}
\def\gsim{\lower.5ex\hbox{\gtsima}}

\def\spose#1{\hbox to 0pt{#1\hss}}
\def\lta{\mathrel{\spose{\lower 3pt\hbox{$\mathchar"218$}}
     \raise 2.0pt\hbox{$\mathchar"13C$}}}
\def\gta{\mathrel{\spose{\lower 3pt\hbox{$\mathchar"218$}}
     \raise 2.0pt\hbox{$\mathchar"13E$}}}

\newcommand{\mincir}{\raise -2.truept\hbox{\rlap{\hbox{$\sim$}}\raise5.truept
\hbox{$<$}\ }}
\newcommand{\magcir}{\raise -2.truept\hbox{\rla669p{\hbox{$\sim$}}\raise5.truept
\hbox{$>$}\ }S}
\newcommand{\minmag}{\raise-2.truept\hbox{\rlap{\hbox{$<$}}\raise 6.truept\hbox
{$>$}\ }}

\newcommand{\be}{\begin{equation}}
\newcommand{\ee}{\end{equation}}
\newcommand{\ba}{\begin{eqnarray}}
\newcommand{\ea}{\end{eqnarray}}
\newcommand{\brr}{\begin{array}}
\newcommand{\err}{\end{array}}
\newcommand{\bc}{\begin{center}}
\newcommand{\ec}{\end{center}}

\def\spose#1{\hbox to 0pt{#1\hss}}
\def\gsim{\mathrel{\spose{\lower 3pt\hbox{$\mathchar"218$}}
          \raise 2.0pt\hbox{$\mathchar"13E$}}}
\def\lsim{\mathrel{\spose{\lower 3pt\hbox{$\mathchar"218$}}
          \raise 2.0pt\hbox{$\mathchar"13C$}}}
\begin{document}

\title[the early X-ray afterglow] {Can the early X-ray afterglow 
of GRBs be explained\\ 
by a contribution from the reverse shock?
}

\author[Genet et al.]{F. Genet, F. Daigne \& 
R. Mochkovitch\\
Institut d'Astrophysique de Paris - UMR 7095 CNRS et Universit\'e
Pierre et Marie Curie, \\
98 bis, boulevard Arago, 75014 Paris, France\\ 
\tt e-mail: mochko@iap.fr}

\maketitle

\begin{abstract}
We propose to explain the recent observations of GRB early X-ray afterglows with SWIFT by the
dissipation of energy in the reverse shock which crosses the ejecta as it is decelerated by the
burst environment. We compute the evolution of the dissipated power and 
discuss the possibility
that a fraction of it can be radiated in the X-ray range. 
We show that this reverse shock contribution 
behaves in a way  very similar to the observed 
X-ray afterglows if the following two conditions are satisfied:
({\it i}) the Lorentz factor of the material which is 
ejected during the late stages 
of source activity decreases to small values $\Gamma<10$ and ({\it ii})
a large part of the shock dissipated energy is transferred to a 
small fraction ($\zeta \lsim$ $10^{-2}$) 
of the electron population. We also
discuss how our results may help to solve 
some puzzling problems raised by multiwavelength early afterglow observations 
such as the presence of chromatic breaks.  
\end{abstract}
\section{Introduction}
The X-Ray Telescope (XRT) on board the SWIFT satellite has for the first time 
allowed a follow-up 
of the X-ray afterglows of GRBs starting within one minute of the BAT 
trigger (Burrows et al. 2005a).
These early afterglow observations 
have revealed several surprising features which cannot be 
easily understood 
in terms of the usual interpretation where the afterglow comes from 
dissipation in a forward
shock propagating through the source environment. At very early times 
immediately after the
burst prompt emission, the afterglow first exhibits a steep 
decrease of temporal slope 
$\alpha_1 \sim 3$  - 5 ($F_{\rm X} \propto t^{-\alpha}$)
(Tagliaferri et al. 2005). It is often followed 
by a much shallower part with $0.2 < \alpha_2 < 0.8$ which can last for 
several hours until a more standard slope $1 < \alpha_3 < 1.5$ is finally
observed (Nousek et al. 2005). Moreover flares with sharp rise and decay times are often present, 
superimposed on the
power-law evolution (Burrows et al. 2005b). In most cases, the spectrum remains essentially 
constant through the breaks
which may indicate that a single physical process is responsible 
for the whole X-ray emission. 
The forward shock could be such 
a process but it seems unable,
at least in its simplest version, to account for the early slopes 
$\alpha_1$ and $\alpha_2$. It has been
suggested that the shallow part of the light curve could still be produced 
by the forward shock if it is continuously feeded in energy by the 
central source (Panaitescu et al. 2005; Zhang et al. 2005). 
Another possibility would be to delay the rise of the forward shock
contribution as a result of viewing angle effects (Eichler and Granot, 2005).
These two proposals  
would however strengthen the constraint on the 
efficiency of the prompt gamma-ray
emission which is already a potential problem for the internal shock scenario
(see however Fan \& Piran, 2006 and Zhang et al. 2006). 
For the initial steep decay,
curvature effects of the emitting shell have been invoked (Nousek et al.
2005) while
flares are usually explained by a late activity of the central
source (Zhang et al. 2005; Fan \& Wei, 2005). 

In this paper we do not consider the origin of flares but rather 
focus on the evolution of the early X-ray afterglow.
We propose that it could be accounted for by a contribution from the 
reverse shock. We develop a simple model which allows us to
follow the internal, reverse and forward shocks in a consistent way.
We compute the energy dissipated in the reverse shock and show that,
for some specific initial distribution of the Lorentz factor in the
flow, it is possible to reproduce the 
succession of the three slopes $\alpha_1$, $\alpha_2$ and $\alpha_3$.
We then discuss under which conditions part of this dissipated energy can be
radiated in the X-ray range, providing an alternative explanation for
the early X-ray afterglows of GRBs. We also obtain the optical 
emission of the reverse shock and show that chromatic breaks can be
observed in some cases. 

The paper is organized as follows: in Sect.2 we present the simplified model
we use to follow the dynamics of internal shocks. We explain in Sect.3 
how it is extended to include the interaction with the environment and
we compute the power dissipated in the reverse shock. We consider in Sect.4
the possibility for this power to be partially radiated in the X-ray range. 
We discuss in Sect.5 the relative importance of the reverse and forward shock 
contributions and present X-ray and optical afterglow light curves
produced by the reverse shock alone.  
Sect.6 is our conclusion.

\section{The origin of GRB pulses}
In the context of the internal shock model for the prompt emission of GRBs
the pulses observed in the light curve are produced when 
fast moving material catches up with slower one previously 
ejected by the central source (Rees \& Meszaros, 1994). 
This process has often
been represented by the collision of two ``shells'' of negligible
thickness. However the central source probably does not release 
individual shells but a continuous relativistic outflow with a
varying Lorentz factor. For this reason the shape of the pulses
is largely 
dominated by hydrodynamical effects (Daigne \& Mochkovitch, 2003) while 
high latitude emission
(curvature effect) only becomes important at late times. 
Soderberg and Fenimore (2001) 
have for example found
that the decay of pulses differs 
from what would be expected if it was controlled by 
the curvature effect alone. 
A hydrodynamical study 
of the relativistic flow emerging from the central engine therefore
appears necessary for a detailed description of the physics of pulses
but it is naturally quite expensive in computing time (Daigne \&
Mochkovitch, 2000).
Fortunately it can often be replaced by a simplified model where the 
flow is represented by a large number of regularly ejected 
shells which interact by
direct collision only (Daigne \& 
Mochkovitch, 1998). This neglects pressure waves but 
this is a good approximation since kinetic energy strongly dominates over
internal energy of the flow. This approach implies to use  
many shells (from $10^3$ to $10^4$) to represent accurately the distribution
of mass and Lorentz factor. It is different 
from the even more simplified description where the numbers of shells
essentially corresponds to the number of pulses to be produced and where
the temporal profiles are then entirely fixed by the curvature effect
(Kobayashi, Piran \& Sari, 1997).
\\
    
The position $R_i$ of each shell of mass $M_i$ and Lorentz factor $\Gamma_i$
is followed as a function of time $t$ (in the source frame). When shell
$i$ catches up with shell $i+1$ a shock occurs at time $t_{\rm s}$ and
radius $R_{s}$. The two shells merge
and the resulting Lorentz factor after the collision is given by
\begin{equation}
\Gamma_{\rm r}=\sqrt{\Gamma_i\Gamma_{i+1}{m_i\Gamma_i+m_{i+1}\Gamma_{i+1}
\over m_i\Gamma_{i+1}+m_{i+1}\Gamma_i}}\ .
\end{equation}
If the the released energy
\begin{equation}
E=m_i\Gamma_i c^2+m_{i+1}\Gamma_{i+1} c^2-(m_i+m_{i+1})\Gamma_{\rm r} c^2
\end{equation}
can be efficiently radiated it will be received by the observer at a time
\begin{equation}
t_{\rm obs}=t_{\rm s}-{R_{\rm s}\over c}
\end{equation}
and for a typical duration 
\begin{equation}
\Delta t_{\rm obs}= {R_{\rm s}\over 2c \Gamma_{\rm r}^2}
\end{equation}
under the condition that the radiative time is 
much smaller than the dynamical time (fast cooling regime). 
The burst bolometric luminosity can then be obtained from the sum of
all the elementary shock contributions, the dynamical evolution 
being terminated
when all the shells have their Lorentz factor decreasing downstream
so that no new internal shock can form.
For an accurate description of the pulse profile at late times,
the contribution $\ell(t)$ of each elementary shock must include 
the curvature effect of the emitting shell  which yields
\begin{equation}
\ell(t)={2E\over \Delta t_{\rm obs}
(1+{t-t_{\rm obs}\over \Delta t_{\rm obs}})^3}
\end{equation}
for $t_{\rm obs}<t<t_{\rm obs}+\left(1-{\rm cos}
\Delta \theta\right)R_{\rm s}/c$, where $\Delta \theta$ is the opening
angle of the jet here supposed to be seen on axis 
(Granot, Piran \& Sari, 1999; Woods \& Loeb, 1999).
The luminosity in a given energy band depends on some additional  
(and uncertain) assumptions on the post-shock magnetic field and Lorentz factor
of the electrons which are discussed in Sect. 4 and 5 while in Sect. 3 
we restrict
ourselves to the bolometric
emission only. 

To produce a single pulse burst (or a pulse as a building block of 
a more complex burst) we have often used in previous works 
(Daigne \& Mochkovitch, 1998, 2000)
an initial distribution of the Lorentz factor of the form
\begin{equation}
\Gamma(t)={\Gamma_{\rm max}+\Gamma_{\rm min}\over 2}
-{\Gamma_{\rm max}-\Gamma_{\rm min}\over 2}\,{\rm cos}\,\Big(\pi {t\over
0.2\,t_{\rm W}}\Big)
\end{equation}
if $t< 0.2\,t_{\rm W}$ and $\Gamma(t)=\Gamma_{\rm max}$ if
$t> 0.2\,t_{\rm W}$; $\Gamma_{\rm max}$ and $\Gamma_{\rm min}$
are the maximum and minimum values of the Lorentz factor and $t_{\rm W}$ the
duration of the relativistic wind emission (the first shell is then ejected at
$t=0$ and the last one at $t=t_{\rm W}$).
This Lorentz factor distribution 
where a rapid part of the flow is decelerated by a slower part placed 
ahead of it,
has been represented in Fig.1a for $\Gamma_{\rm max}=200$,
$\Gamma_{\rm min}=50$ and $t_{\rm W}=10$ s.
The resulting bolometric
profile from dissipation by internal shocks is shown
in Fig.1b for a total (isotropic) radiated energy
$E_{\rm rad}=10^{53}$ erg.
\begin{figure*}{}
\begin{center}
\begin{tabular}{cc}
\resizebox{0.49\hsize}{!}{\includegraphics{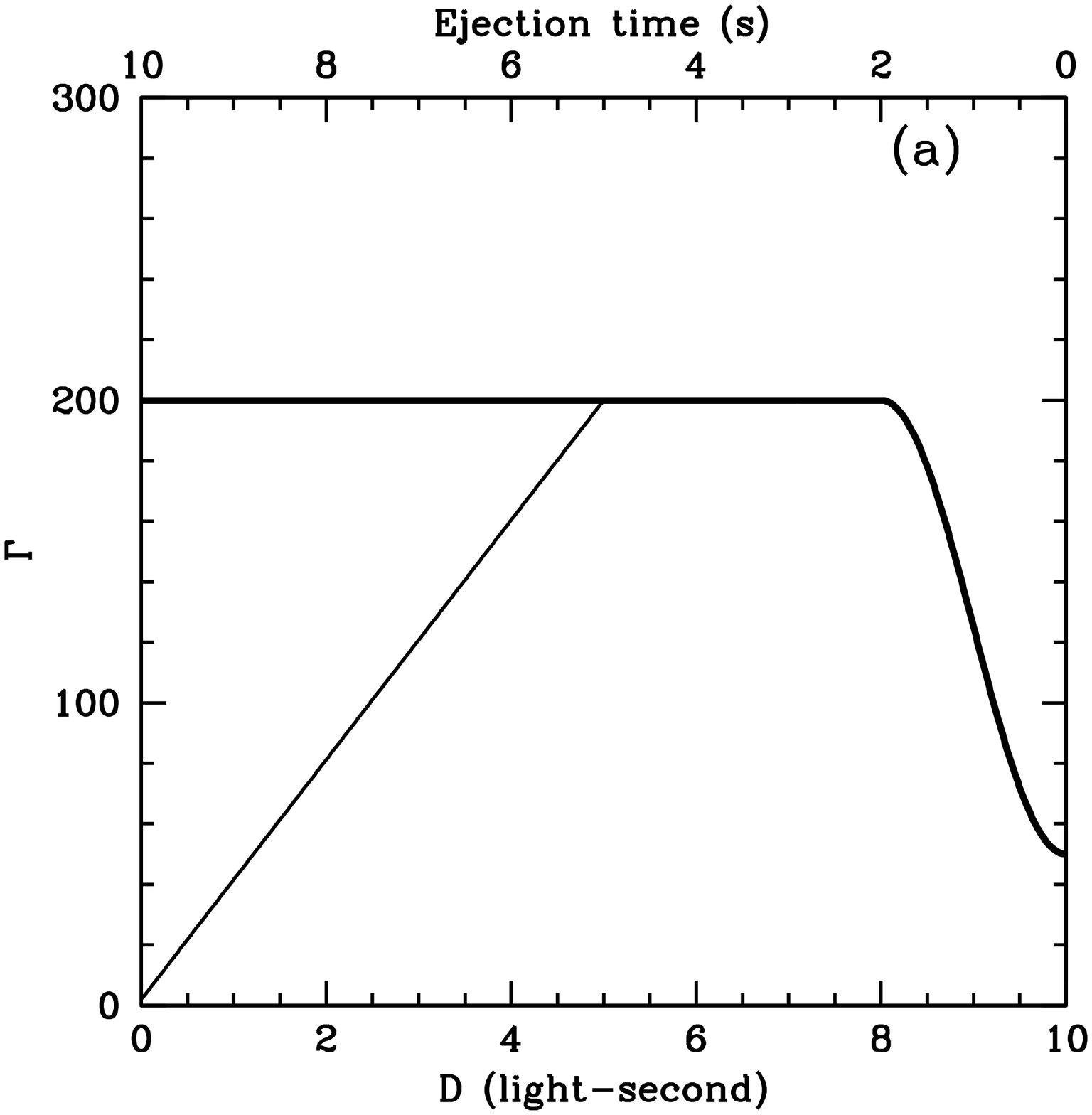}}&
\resizebox{0.49\hsize}{!}{\includegraphics{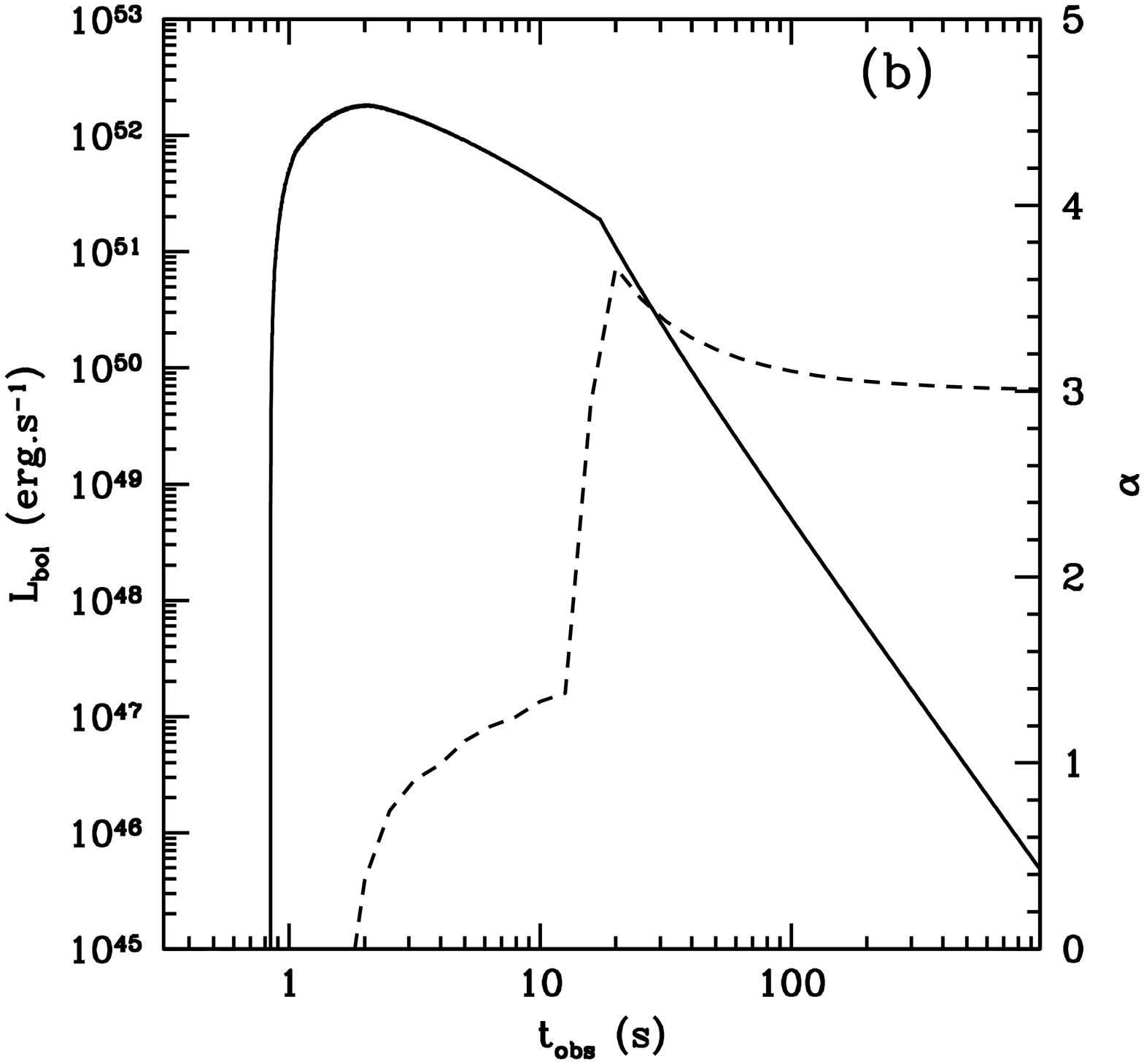}}\\
\end{tabular}
\end{center}
\caption{A single pulse burst. (a): Initial distribution of the Lorentz factor in the relativistic
flow as a function of ejection time and distance $D$ to the source (in 
light-seconds) according to Eq.(6) with $\Gamma_{\rm max}=200$,
$\Gamma_{\rm min}=50$ and $t_{\rm W}=10$ s (thick full line) and 
to Eq.(7)
with $\delta=1$ and $\Gamma_{\rm f}=2 $ (thin full line);
(b): Bolometric profile for the distribution given by Eq.(6) (full line)
together
with the corresponding temporal slope $\alpha$ (dashed line). 
After maximum, the profile is first controlled by the dynamics of 
internal shocks 
before the curvature effect eventually dominates after 
$t_{\rm obs}\sim 20$ s.} 
\end{figure*}

The decline of the pulse after maximum is 
first controlled by the dynamics of internal shocks.
This 
would lead to an asymptotic behavior  
$L(t)\propto t^{-3/2}$ if it was not  interrupted at a time
$\tau\sim t_{\rm W}$ 
when all the fast material of the ejecta has been shocked.
Daigne and Mochkovitch (2003) have shown that the gamma-ray profiles
which can be obtained from this first part of the bolometric 
light curve are gene\-ral\-ly 
in good agreement with the 
early decline following maximum count rate in observed GRBs
(Ryde \& Svensson, 2000). 

After all the ejecta has been shocked the pulse evolution 
becomes fixed by geometry, the contribution of each 
shocked shell being given by Eq.(5).
At large times $t\gg \tau$, 
all the $\ell(t)$ and therefore the global profile $L(t)$ 
asymptotically behave as $t^{-3}$. 
However at 
early times $t\gsim \tau$,
a steeper decline can be obtained (Nousek et al. 2005) as 
illustrated in Fig.1b where the temporal slope $\alpha$
has been plotted together with the profile.
It has a maximum of 3.65 just at the end of the
internal shock phase before relaxing to 3 after a few $\tau$.
 
\section{The reverse shock}
\subsection{Physical description}
The profile calculated above
corresponds to a ``naked GRB'' (Kumar \& Panaitescu, 2000) and would be the only component observed
in the absence of external medium. The burst environment will however 
interact with the ejecta, leading to a forward shock propagating
through the circumstellar medium  
and a reverse shock sweeping back into the ejecta.
We compute below (Sect.3.2) the power dissipated in the reverse
shock and discuss its possible contribution to the early X-ray emission
of GRBs. With the initial distribution of the Lorentz factor given by Eq.(6)
the reverse shock crosses the ejecta in a short time and 
cannot explain an emission
lasting for several days.
The situation is however very different if a slight change is made in the 
initial distribution of the Lorentz factor. 
We expect that the central source will not stop ejecting
relativistic material abruptly at $t=t_{\rm W}$.
We instead propose that $\Gamma$ will progressively decrease 
until it reaches  
a small value (possibly close to unity) at $t_{\rm W}$. Since $\Gamma(t)$
is given by the ratio $\dot E/\dot M$ of the energy to mass injection rates,
a small $\Gamma$ can be a consequence of ({\it i}) a decrease of $\dot E$,
less and less energy 
becoming available from the source
to accelerate a given baryon load or/and ({\it ii}) a catastrophic increase of 
$\dot M$. Case ({\it i}) appears more natural during the
late stages of source activity and has been adopted 
in presenting our results in Sect.3.2.

We have then introduced
a new distribution of the Lorentz factor where, for $t>0.5\,t_{\rm W}$,
$\Gamma(t)$ decreases to a final value 
$\Gamma_{\rm f}$  
\begin{equation}
\Gamma(t)=\Gamma_{\rm f}
+\left({\Gamma_{\rm max}-\Gamma_{\rm f}}\right)\,
\left[{1-t/t_{\rm W}\over 0.5}\right]^{\delta}
\end{equation} 
while for $t<0.5\,t_{\rm W}$, $\Gamma(t)$ is still given by Eq.(6).
This mo\-di\-fied Lorentz factor is plotted in Fig.1a for 
$\Gamma_{\rm max}=200$, $\Gamma_{\rm f}=2$, $t_{\rm W}=10$ s 
and $\delta=1$. 
With this new distribution, 
the duration of source activity remains unchanged but
the reverse shock will be present 
for a much longer time, until all the ejecta has been decelerated to 
$\Gamma\sim \Gamma_{\rm f}$. 
The forward shock also remains feeded in energy
as slow material from the ejecta is continuously catching up but
the resulting effect is too small in this case to account 
for the shallow part of the 
light curve (we assumed that equal amounts of kinetic energy
are injected before and after $t=0.5\,t_{\rm W}$). 

To compute the energy dissipated in the reverse shock
we had to implement
in our shell model the interaction with the burst environment.
This was done by considering the contact
discontinuity which separates the ejecta and the shocked
external medium. In our simple description it is represented
by two shells moving at the same Lorentz factor $\Gamma$.
The first one corresponds to the mass $M_{\rm ej}$ of the
ejecta already crossed by the reverse shock, which
carries a total e\-ner\-gy $\Gamma M_{\rm ej} c^2$,
and the second
to the shocked external medium of mass $M_{\rm ex}$.  If the forward
shock moves quasi-adiabatically (slow cooling regime), 
this shell
keeps its internal energy (since $pdV$ work is
neglected in our simple model) so that its
total energy is $\Gamma \Gamma_i M_{\rm ex} c^2$
where $(\Gamma_i-1)c^2$ 
is the dissipated energy (per unit mass) in 
the fluid rest frame.  
Two processes will affect this two shell structure 
at the contact discontinuity : it will collide either with shells of the
external medium at rest, or with rapid shells of the relativistic ejecta
catching up. 
This represents both the forward and reverse shock in our simplified 
picture.
\vskip 0.3cm\noindent
\textit{Forward shock:}
the interaction with
the external medium is discretized by assuming that
a collision occurs each time
the contact discontinuity has travelled from a radius $R$
to a radius $R^{\,\prime}$ so that the swept-up mass is
\begin{equation}
m_{\rm ex}=\int_R^{R^{\,\prime}} 4\pi r^2 \rho(r) dr=q\,{M\over
\Gamma}
\end{equation}
where $M=M_{\rm ej}+M_{\rm ex}$, $\rho(r)$ is the density of the external
medium (for which we
adopted either a constant or a stellar wind distribution) and
$q\ll 1$ (we take in practice
$q=10^{-2}$). Writing the conservation of energy-momentum for this collision, 
we obtain the new Lorentz factor $\Gamma_r$ 
at the contact discontinuity 
\begin{equation}
\Gamma_r=\left[{(M_{\rm ej}+M_{\rm ex}\Gamma_i)\Gamma^2+m_{\rm ex}\Gamma\over
(M_{\rm ej}+M_{\rm ex}\Gamma_i)+2 m_{\rm ex}\Gamma}\right]^{1/2}
\end{equation}
and also the new Lorentz factor $\Gamma_i^{\prime}$ 
for internal motions after the collision
\begin{equation}
\Gamma_i^{\prime}={(M_{\rm ej}+M_{\rm ex} \Gamma_i)\Gamma+m_{\rm ex}-M_{\rm ej}\Gamma_r\over
(M_{\rm ex}+m_{\rm ex})\Gamma_r}\ .
\end{equation}
It should be noted that the above equations assume  
that material 
in the burst environment is at rest. This neglects the pair-loading
process resulting from the initial flash of gamma-rays 
which pre-accelerates the circumstellar medium (Madau \& Thompson, 2000;
Thompson \& Madau, 2000; Beloborodov, 2002) out to a radius
\begin{equation}
R_{\rm acc}\sim 7\,10^{15}\,E_{\gamma,\,53}^{1/2}\ \ {\rm cm}
\end{equation}
where $E_{\gamma,\,53}$ is the isotropic gamma-ray energy of the flash in units of
$10^{53}$ erg. For this reason, the deceleration 
by the external medium will be delayed by
\begin{equation}
\Delta t_{\rm acc}\sim {R_{\rm acc}\over 2c \Gamma^2}=
12\,E_{\gamma,\,53}^{1/2}\,\Gamma_2^{-2}\ \ {\rm s}
\end{equation}
where $\Gamma_2$ is the average Lorentz factor of the ejecta in units of
$10^2$. 
Therefore the initial dynamical evolution will be that
of a naked GRB but the effect will last more than one minute only for the 
most extreme bursts with $E_{\gamma,\,53}>10$.    \\ 

\noindent
\textit{Reverse shock:}
as the Lorentz factor at the contact discontinuity decreases,
new shells from the ejecta become able to catch up.
Writing again the conservation of energy-momentum for these collisions,
we obtain the
change in Lorentz factor 
\begin{equation}
\Gamma_r=\sqrt{\Gamma\gamma_{\rm ej}}\left[{(M_{\rm ej}+M_{\rm ex}\Gamma_i)\Gamma+m_{\rm ej}\gamma_{\rm ej}
\over (M_{\rm ej}+M_{\rm ex}\Gamma_i)\gamma_{\rm ej}+m_{\rm ej}\Gamma}\right]^{1/2}
\end{equation}
and the related dissipated energy
\begin{equation}
E_{\rm diss}=(M_{\rm ej}+M_{\rm ex}\Gamma_i)\Gamma c^2+
m_{\rm ej}\gamma_{\rm ej} c^2
-(M_{\rm ej}+m_{\rm ej}+M_{\rm ex} \Gamma_i)\Gamma_r c^2
\end{equation}
where $m_{\rm ej}$ and $\gamma_{\rm ej}$ are respectively the mass and Lorentz factor of the
colliding material from the ejecta.
\subsection{The dissipated power}
Using this simplified model for the interaction of the ejecta with
its environment we can describe the deceleration of the front shell and the
propagation of the reverse shock. We have obtained the dissipated power in the 
reverse shock for different burst environments (uniform medium or wind).
In the wind case, we considered three values of the
parameter $A_*$: 0.5, 0.1 and 0.05 (such as 
$\rho(r)=5\,10^{11}\,A_*/r^2$ g.cm$^{-3}$ with $A_*=1$ 
for a wind mass loss
rate $\dot M_{\rm w}=10^{-5}$ M$_\odot$.yr$^{-1}$ and a terminal velocity 
$v_{\infty}=1000$ km.s$^{-1}$). In the constant density case, we 
also tried three values of $n$: 1000, 10 and 0.1 cm$^{-3}$.
The resulting profiles are shown in Fig.2 for $\Gamma_{\rm f}=2$
and $\delta=1$ in Eq.(7) but we checked that they remain
essentially unchanged when $\Gamma_{\rm f}$ is varied between 1 and 10
and $\delta$ between 0.5 and 2. 

\begin{figure*}{}
\begin{center}
\begin{tabular}{cc}
\resizebox{0.49\hsize}{!}{\includegraphics{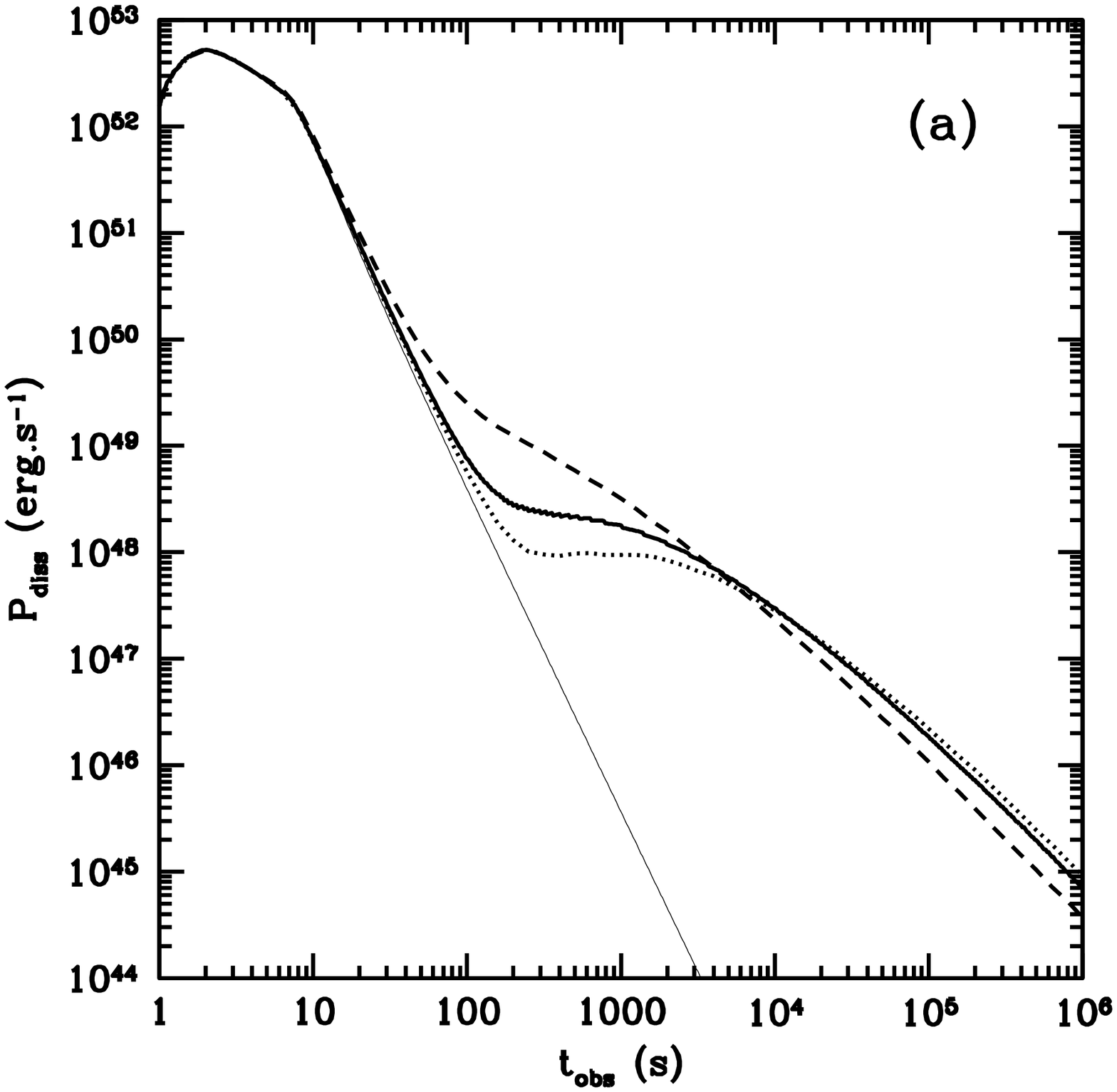}}&
\resizebox{0.49\hsize}{!}{\includegraphics{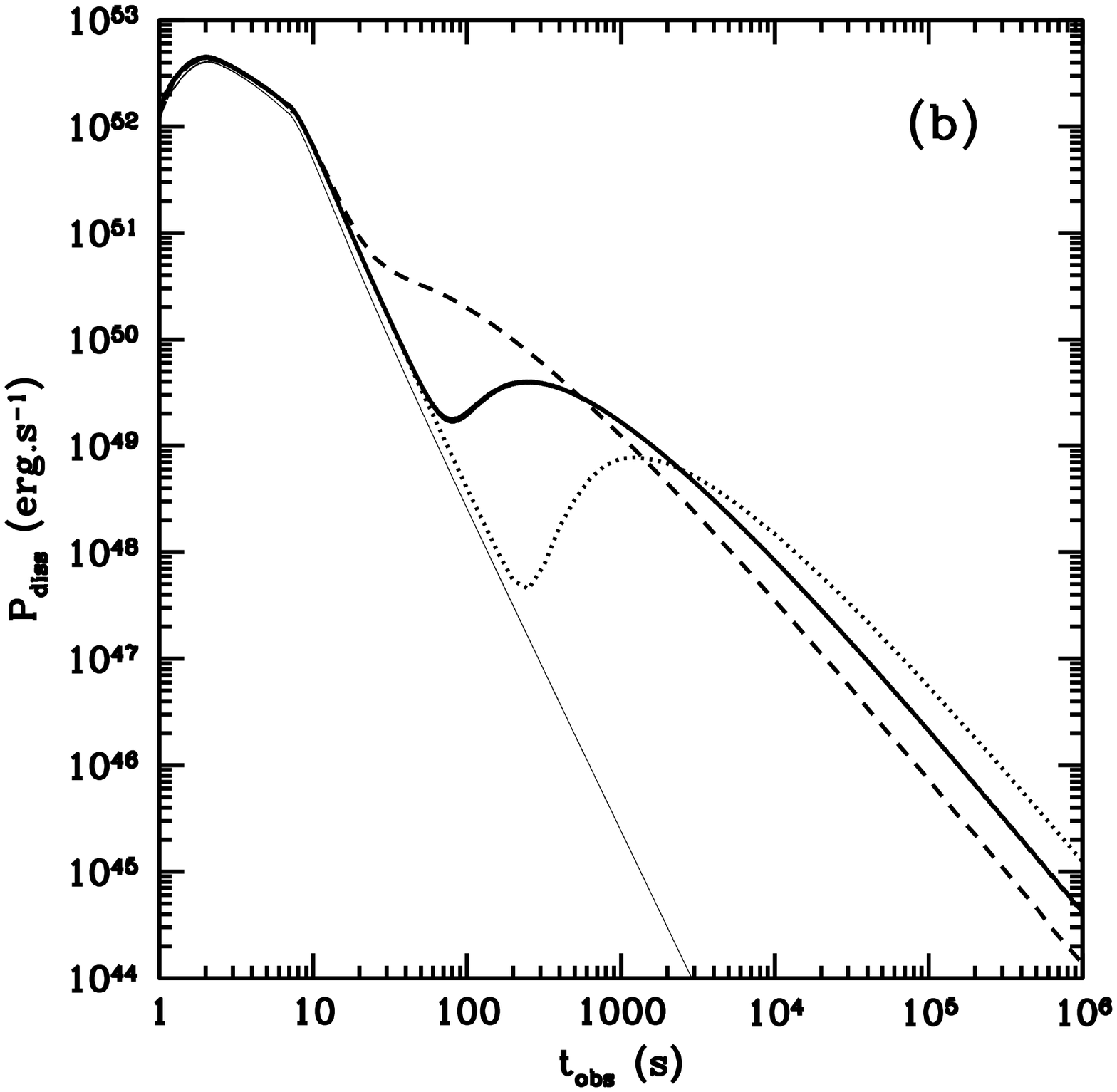}}\\
\end{tabular}
\end{center}
\caption{Dissipated power as a function of observer time during
reverse shock propagation when the Lorentz factor is given
by Eq.(7) with $\Gamma_{\rm f}=2$ and $\delta=1$. 
(a): wind case with $A_*=0.5$ (dashed line),
$A_*=0.1$ (full  line) and $A_*=0.05$ (dotted line); 
(b): uniform density case with
$n=1000$ (dashed line), 10 (full line) and 0.1 cm$^{-3}$ (dotted line). 
In each panel the thin full line represents the naked burst.  
}
\end{figure*}

The curves in Fig.2 show a striking
resemblance with the early X-ray afterglows observed by SWIFT.
After about 100 s the 
reverse shock component 
dominates over the tail of the 
of the burst prompt emission
computed in the last section.  
At late times the decline follows a constant 
slope $\alpha\sim 1.5$. The shape of the intermediate region 
is most sensitive to the density of the 
burst environment. At high density it is nearly suppressed, the constant 
slope $\alpha \sim 1.5$ following directly the initial steep
decrease. Conversely at low density, it
can become completely flat and even fall to a temporary minimum.
For comparison, we have also represented in Fig.2 the profile 
corresponding to the 
naked GRB. Without an external medium and
for the distribution of Lorentz factor given by Eq.(7)
(with $\Gamma_{\rm f}=2$ and $\delta=1$) we find that a large fraction 
of the ejecta (all slow material with $\Gamma<140$)  
remains unaffected by internal shocks. 
With an external medium the reverse shock 
propagates throught this material which produces the additional power
at late times. 
\\

To elucidate the behavior of the reverse shock contribution
we have considered the following 
simplified case which can be handled analytically: 
the ejecta is supposed to be made of a rapid single 
shell of mass $M_0$ and initial 
Lorentz factor $\Gamma_0$ (representing the fast material
where the prompt emission takes place) followed by a slower tail
of the form
\begin{equation}
\Gamma(M)=\Gamma_{\rm f}+(\Gamma_0-\Gamma_{\rm f})\left({M\over M_{\rm s}}
\right)^
{\delta}
\end{equation}
where $\Gamma_{\rm f}$ is the final Lorentz factor at the end of the tail
and $M_{\rm s}$ is the 
total mass of the slow material
($M=0$ corresponds to the last 
emitted shell and $\delta$ allows to vary the tail shape). Notice
that this expression of $\Gamma(M)$ directly results from Eq.(7)
if $\dot M$ is constant.
Such a distribution of $\Gamma$ skips the prompt phase  
and the resulting dissipated power $P_{\rm diss}$ 
comes from the 
reverse shock only.

The reverse shock contribution is maximum at a time 
close to the deceleration time of the front shell 
\begin{equation}
t_{\rm dec}={R_{\rm dec}\over 2 c\,\Gamma_0^2}\ \ \ 
{\rm with}\ \ \ R_{\rm dec}=\left[{M_0\,(3-s)\over 4\pi A \Gamma_0}
\right]^{1\over 3-s}
\end{equation}
where we have used the notation $\rho=A r^{-s}$ with $A=\rho$ and
$s=0$ for a uniform medium and $s=2$ for a stellar wind.
A full analytical solution can be obtained for the reverse shock 
contribution but we only derive below its
asymptotic behavior 
assuming that the front shell 
essentially follows the Blandford-McKee solution, i.e. it is
only weakly affected by the additional energy coming from the slow material
progressively catching up. We have checked this approximation with 
the numerical simulations and it is satisfied to an accuracy of about
25\%. 

We can write the power dissipated in the
reverse shock as
\begin{equation}
P_{\rm diss}=- {dM\over d\Gamma}{d\Gamma\over dt}\,
\Gamma\, e c^2
\end{equation} 
where $t$ is the observer time when a shell of Lorentz factor
$\Gamma$ catches up with the front shell. 
The fraction $e$ of the in\-co\-ming material 
kinetic energy
dissipated in the collision can be obtained from Eq.(13) and (14) 
with $m_{\rm ej}\ll M_{\rm ej}+M_{\rm ex}\Gamma_i$, leading to
\begin{equation}
e={1\over 2}\left[1-\left({\Gamma_{\rm fs}\over \Gamma}\right)\right]^2
\end{equation}
where $\Gamma_{\rm fs}$ is the Lorentz factor of the front shell given
by the Blandford-McKee solution
\begin{equation}
\Gamma_{\rm fs}\simeq \Gamma_0\left({R\over R_{\rm dec}}\right)^{-\lambda}
\end{equation}
with $\lambda={3-s\over 2}=3/2$ (resp. 
$1/2$) for a uniform medium (resp. a stellar wind). Using 
$dR/dt=2c\,\Gamma_{\rm fs}^2$ we then get the relation between shock radius and
observer time
\begin{equation}
{t\over t_{\rm dec}}={1\over 2\lambda +1}\left({R\over R_{\rm dec}}\right)
^{2\lambda+1}\ .
\end{equation} 
With our assumed distribution (Eq.(15)) of the  Lorentz factor in the 
slow material which 
is steadily increasing outwards, each shell moves independently
at the constant Lorentz factor $\Gamma$ until it catches up with 
the forward shock.
We moreover neglect the fact that the slow material is emitted over 
a certain duration and we write the position of each shell as a function 
of observer time as
\begin{equation}
{R\over R_{\rm dec}}={t\over t_{\rm dec}}\left({\Gamma\over \Gamma_0}\right)^2
\ .
\end{equation} 
Notice that if the Lorentz factor is not initially monotonic in the
slow material, internal shocks will take place which, when they are completed,
will leave a new distribution of $\Gamma$, now steadily increasing outwards.
For observing times long compared to the time of internal shocks
Eq.(21) will therefore still hold.
 
Eliminating the radius between Eq.(20) and (21) gives the time when
a shell of Lorentz factor $\Gamma$ catches up with the forward shock
\begin{equation}
{t\over t_{\rm dec}}=(2\lambda+1)^{1/2\lambda}
\left({\Gamma\over \Gamma_0}\right)^{-{2\lambda+1\over \lambda}}\ .
\end{equation} 
Now from Eq.(19), (20) and (22) we can obtain the Lorentz factor $\Gamma_{\rm fs}$
of the forward shock when the slow shell of Lorentz factor $\Gamma$
catches up
\begin{equation}
\Gamma_{\rm fs}=\Gamma\,(2\lambda+1)^{-1/2}\  .
\end{equation} 
To end with a simple power law expression for the dissipated power 
we write 
\begin{equation}
{dM\over d\Gamma}={M_{\rm s}\over \delta\, \Gamma_0}
\left({\Gamma\over \Gamma_0}\right)^{{1-\delta\over \delta}}
\end{equation}
i.e. we use Eq.(15) with $\Gamma_{\rm f}=0$ which is obviously
uncorrect but does not change the behavior of the solution 
in the re\-la\-tivistic regime. 
Using Eq.(22), (23) and (24) we finally get
\begin{equation}
P_{\rm diss}(t)=\Phi(\lambda,\delta)
\,{\Gamma_0 M_{\rm s}c^2\over \delta\,t_{\rm dec}}
\left({t\over t_{\rm dec}}\right)^{-\left[{3\lambda+1+\lambda/\delta\over 
(2\lambda+1)}\right]}
\end{equation}	  
where 
\begin{equation}
\Phi(\lambda,\delta)={\lambda\over 2}\,
\left[1-\left(2\lambda+1\right)^{-1/2}\right]^2
\times \left(2\lambda+1\right)^{{1-\delta(1+4\lambda)\over 
2\delta(2\lambda+1)}}\ .
\end{equation}
For $\delta=1$ and the two values of interest for $\lambda$, Eq.(25) becomes
\begin{eqnarray}
{P_{\rm diss}\over {\Gamma_0 M_{\rm s}c^2/t_{\rm dec}}}& = &
6.6\,10^{-2} \left({t\over t_{\rm dec}}\right)^{-7/4}\ \ (\lambda=3/2)\nonumber\\ 
& = & 1.5\,10^{-2}\left({t\over t_{\rm dec}}\right)^{-3/2}\ \ (\lambda=1/2)
\end{eqnarray}
which is in good agreement with the asymptotic behavior of the light curves 
shown in  Fig.2.
\begin{figure*}
\begin{center}
\begin{tabular}{cc}
\resizebox{0.49\hsize}{!}{\includegraphics{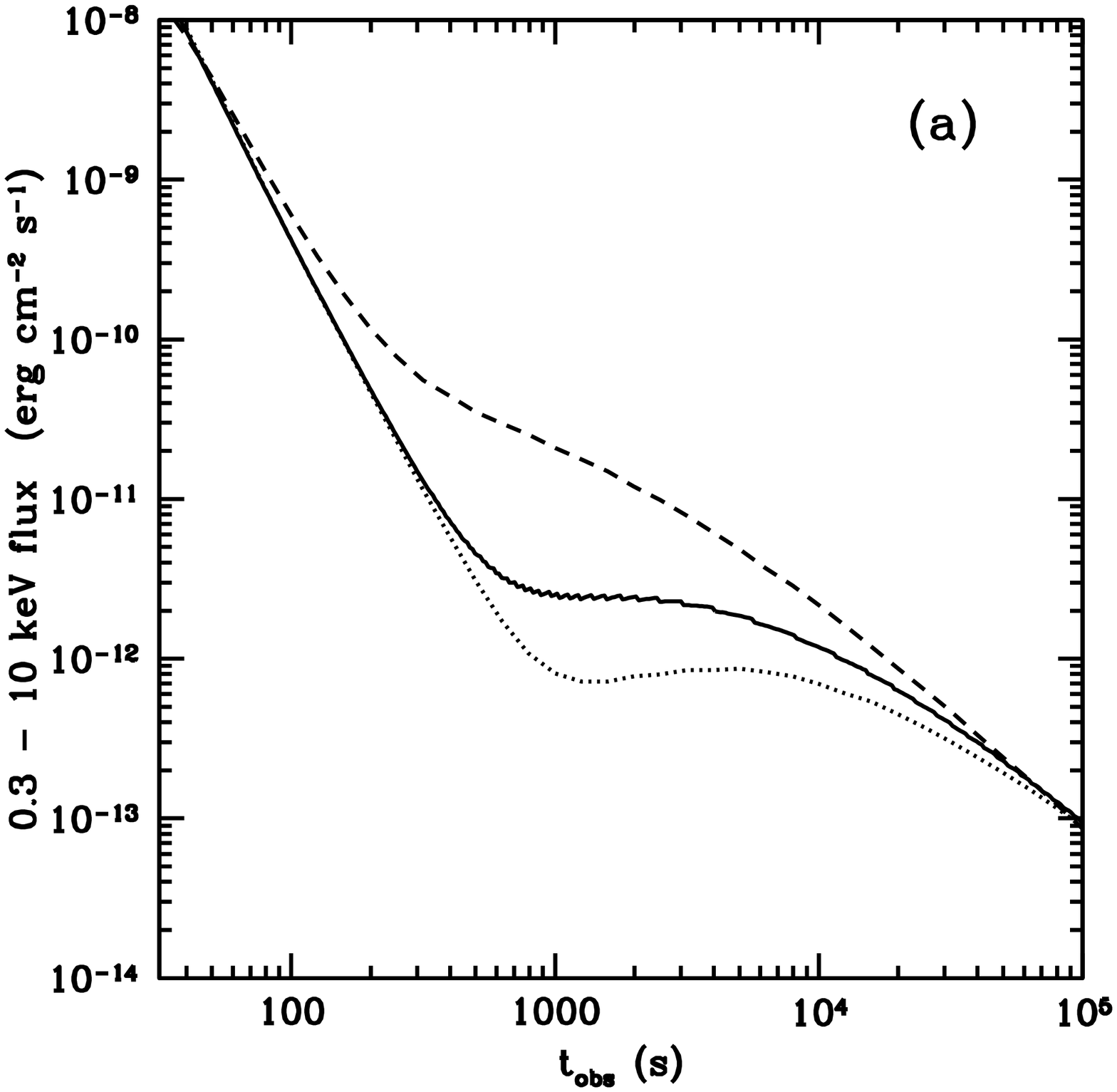}}&
\resizebox{0.49\hsize}{!}{\includegraphics{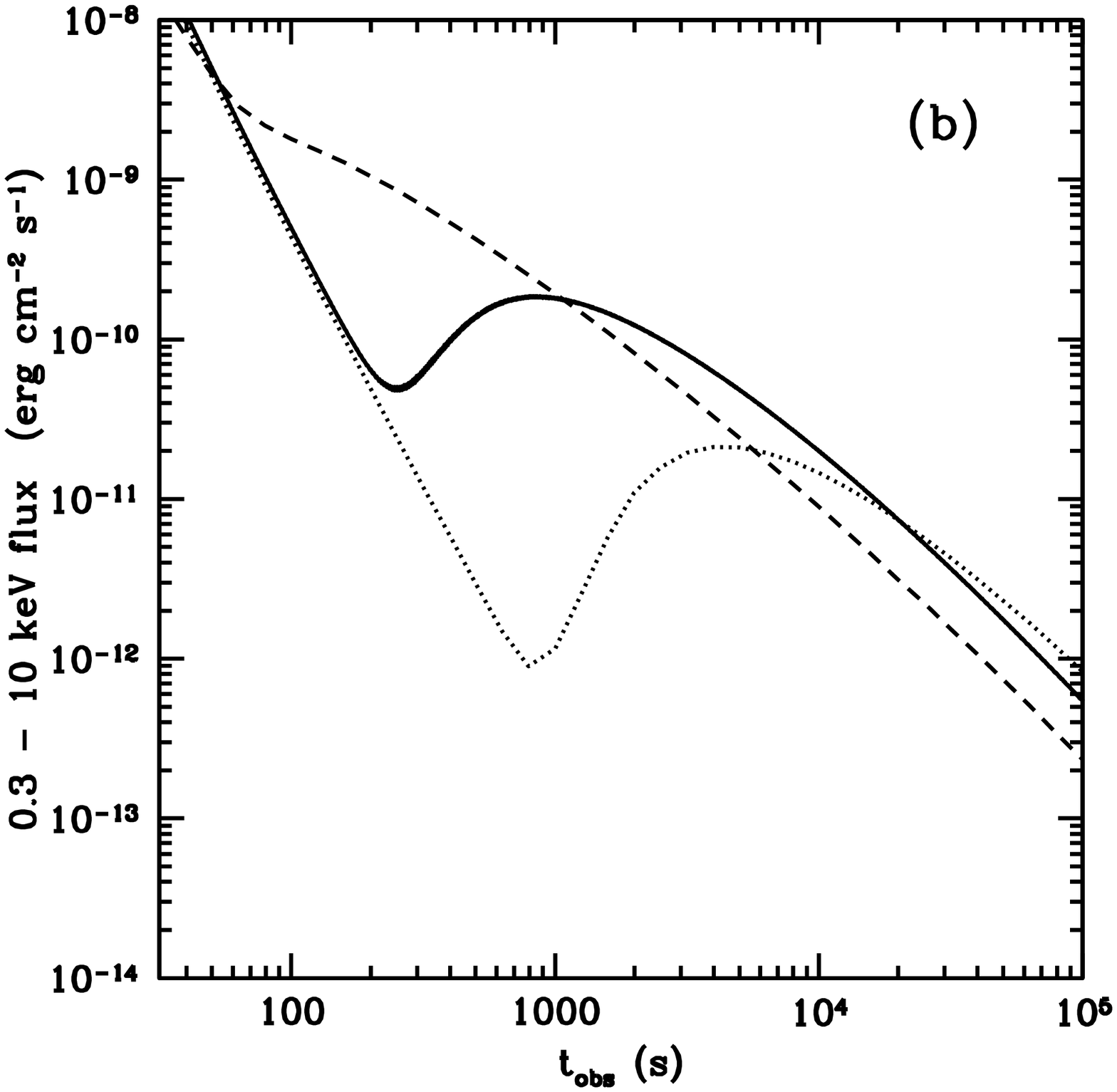}}\\
\end{tabular}
\end{center}
\caption{Synthetic X-ray light curves in the 0.3 - 10 keV range
for the models of Fig.2. The adopted post-shock energy redistribution
parameters are $\epsilon_e=\epsilon_B=1/3$ and $\zeta=10^{-2}$. The
assumed burst redshift is $z=2$; The different cases considered  
in Fig.2 are represented by the same full, dotted and dashed lines.}
\end{figure*}
\section{Can the reverse shock contribute in X-rays?}
Despite their similarity with the SWIFT observations, 
it must remain clear that the profiles shown in Fig.2 only trace
the power dissipated in the reverse shock. With the assumptions
ordinary made to compute the reverse shock contribution in GRBs
it should manifest itself mainly in the visible/IR range (Sari \& Piran, 1999). 
Moreover most of
the emission would generally take place in the slow cooling regime so that the
observed light curve will not necessarily trace the instantaneous energy release.

We therefore investigated whether, under some specific conditions, 
a substantial fraction of the dissipated power can be
({\it i}) radiated in the X-ray range and ({\it ii}) in the fast cooling regime.
If the reverse shock contribution originates from synchrotron radiation of
shock accelerated electrons, the characteristic 
synchrotron energy and cooling time behave as
\begin{equation}
E_{\rm s}\propto B \Gamma_e^2\ \ \ \ {\rm and}\ \ \ \ 
t_{\rm s}\propto B^{-2}\Gamma_e^{-1}
\end{equation}
in the rest frame of the shocked material.
Both the post-shock
magnetic field $B$ and typical electron Lorentz factor $\Gamma_e$ have 
therefore to be large to produce an emission at high energy and on a
short time scale $t_{\rm s}<t_{\rm dyn}$. An estimate of $\Gamma_e$ is usually obtained assuming that 
a fraction $\epsilon_e$ of the dissipated energy is injected into a 
fraction $\zeta$ of the electrons so that
\begin{equation}
\Gamma_e\simeq {\epsilon_e\over \zeta}{m_p\over m_e}\,e
\end{equation}
where $m_p$ and $m_e$ are the proton and electron masses and $e c^2$ is the 
energy dissipated per unit mass in the comoving frame. Similarly the 
post-shock magnetic field can be expressed as 
\begin{equation}
B=\left(8\pi\epsilon_B\,\rho\,e c^2 \right)^{1/2}
\end{equation}
where $\rho$ is the comoving density and $\epsilon_B$ the fraction of the 
dissipated energy tranferred to the magnetic field. 
To have large 
$B$ and $\Gamma_e$ values behind the shock we first 
supposed that a complete equipartion is
established between the electronic, magnetic and baryonic components
so that $\epsilon_e=\epsilon_B=\epsilon_{\rm baryon}=1/3$. 
We also assumed that only a small fraction $\zeta\lsim 10^{-2}$
of the electron population is accelerated in the shock. Adopting
$\zeta= 10^{-2}$ increases $\Gamma_e$ 
by a factor of 100 and hence $E_{\rm s}$ by a factor 
$10^4$ and decreases $t_{\rm s}$ by $10^2$
compared to the standard $\zeta=1$ case.

The possibility to have only a small fraction of electrons being 
accelerated has already been considered by Bykov \& Meszaros (1996) 
and also by Eichler \&
Waxman (2005) in the context 
of GRB afterglows. They showed that $\zeta$ is not well constrained by
the observations and, even if $\zeta\sim 1$ appears slightly favored, 
they included the whole interval $m_e/m_p<\zeta<1$ in their analysis.
In internal shocks, which are very similar to the reverse shock (both
take place in the burst ejecta and are mildly relativistic)
a large $\epsilon_e$ is required to maintain a reasonable global efficiency 
since the
fraction of the total energy dissipated by internal shocks hardly exceeds 10\%. 
A small $\zeta$ is also favored to insure that the emission takes place
in the gamma-ray range as shown by Daigne and Mochkovitch (1998) and
more recently by Lee et al. (2005) in the context of the short hard burst 
GRB 050509b.

Examples of synthetic light curves in the XRT band 0.3 - 10 keV are 
shown in Fig.3 for the cases already consi\-de\-red in Fig.2. They have 
been obtained with $\epsilon_e=\epsilon_B=1/3$, $\zeta=10^{-2}$, 
a slope $p=2.5$ for the electron energy distribution and an assumed
redshift $z=2$, typical of the SWIFT burst population. Especially in the
wind case, they seem able to reproduce many of the observed XRT light
curves. Conversely in the uniform density case we often obtain a depressed minimum
followed by a bump rather than a continuous shallow evolution 

\section{Discussion}
Our proposal to explain the early X-ray afterglow of GRBs 
by a contribution of the reverse shock relies on three well 
defined assumptions: ({\it i}) the Lorentz factor of the material  
ejected at late times by the source has to decrease 
to small values, $\Gamma_{\rm f} < 10$;
({\it ii}) the shock dissipated energy must be transferred to only a small
fraction of the electron population; and ({\it iii})  
the forward shock contribution should lie
below that of the reverse shock, at least during the first hours following burst
trigger.

This last condition requires 
an ineffective transfer of energy to electrons 
($\epsilon_e\lsim 10^{-2}$) or/and ma\-gne\-tic field 
($\epsilon_B\lsim 10^{-5}$) in the material crossed by the 
forward shock. Difficulties to produce a sufficiently large
ma\-gne\-tic field extending over the emitting region of GRB afterglows has
for example been recently emphasized by Milosav\-ljevi\'c \& Nakar (2006).
Then, if the reverse shock dominates in X-rays, what is the
situation in the visible? We have checked that in most cases, 
taking small values of $\epsilon_e$ or/and $\epsilon_B$ in the forward 
shock, equally implies that the reverse shock 
dominates in the visible.  
The consistency of our proposal must therefore be checked not only
with X-ray observations
but also at lower wavelengths.

To better understand the multiwavelength behavior of the reverse 
shock contribution, we have computed the peak flux $F_{\rm max}$
and the characteristic synchrotron and cooling frequencies $\nu_m$
and $\nu_c$ (Sari, Piran \& Narayan, 1998). These three quantities
depend on $t$ (observer time), $N_e$ (total number of shock 
accelerated electrons), $B$ (magnetic field in shocked material),
$\Gamma_e$ (typical electron Lorentz factor) and $\Gamma$ (Lorentz factor 
of the emitting material) in the following way  
\begin{equation}
\begin{array}{l}
F_{\rm max}\propto \Gamma B N_e \\
\nu_m\propto \Gamma\, \Gamma_e^2 B \\
\nu_c\propto \Gamma^{-1} B^{-3} t^{-2}
\end{array}
\end{equation} 
We consider their temporal evolution in the asymptotic regime already
described in Sect.3.2, assuming a wind environment. 
The evolution of $\Gamma$ is given by Eq.(22) which can be
reexpressed as 
\begin{equation}
\Gamma(t)=\Gamma_0 \left({t\over 2\,t_{\rm dec}}\right)^{-1/4}
=119\  \Gamma_2 \left({t\over t_{\rm dec}}\right)^{-1/4}
\end{equation}
where $\Gamma_2=\Gamma_0/100$. 
Eq.(18), (23) and (29) show that $\Gamma_e$ reaches a constant value
\begin{equation}
\Gamma_e=4.3\,10^{-2}\,{m_p\over m_e}\,{\epsilon_e\over \zeta}\,
{p-2\over p-1}=79\,{\epsilon_e\over \zeta}\,{p-2\over p-1} 
\end{equation}
where we have added the normalizing factor ${p-2\over p-1}$ ($p$
being the slope of the relativistic electron distribution) which was not 
present in Eq.(29). For the magnetic field, instead of Eq.(30) it is
easier to use the continuity of the energy density at the contact
discontinuity which yields
\begin{equation}
B=(32\pi \epsilon_B c^2 A)^{1/2}{\Gamma\over R}
\end{equation}
where $A$ is the wind constant such as $\rho(R)=A/R^2$ ($A=5\,10^{11}A_*$).
With Eq.(20) for $R$ and Eq.(32) for $\Gamma$ we get
\begin{equation}
B(t)=3\,10^4\;{\left(\epsilon_B\,A_*\right)^{1/2}
\over t_{\rm dec}\,\Gamma_2} 
\left({t\over t_{\rm dec}}\right)^{-3/4}\ {\rm G}
\end{equation}
Finally, the number of accelerated electrons can be obtained from
Eq.(24) which, for $\delta=1$, gives 
\begin{equation}
N_e(t)={2\,\zeta\,E_{\rm s}\over \Gamma_0\,m_p c^2}\left[1-1.19\,\left({t\over t_{\rm dec}}\right)^{-1/4}
\right]
\end{equation}
where $E_{\rm s}={1\over 2}\Gamma_0\,M_{\rm s} c^2$ is the total energy in the 
slow material.
From Eq.(32), (33), (35) and (36) the expressions for    
$F_{\rm max}$, $\nu_m$ and $\nu_c$ can be computed
\begin{equation}
\begin{array}{l}
F_{\rm max}=1.4\,10^8\,{(1+z)\over D_{28}^2}\,
{(\zeta E_{53})\left(\epsilon_B\,A_*\right)^{1/2}
\over \Gamma_2}\times {1\over t}\ \ {\rm mJ} \\
\nu_m=9.15\,10^{16}\,(\epsilon_B\,A_*)^{1/2}
\left({\epsilon_e\over \zeta}\right)^2\,\left({p-2\over p-1}\right)^2 
\times {1\over t}\ \ {\rm Hz}\\
\nu_c=8.2\,10^8\,{t_{\rm dec}^{1/2}\Gamma_2^2\over (\epsilon_B\,A_*)^{3/2}}
\times t^{1/2}\ \ {\rm Hz}
\end{array}
\end{equation} 
with $E_{53}=E_{\rm s}/10^{53}$ erg and where the expression for $F_{\rm max}$
has been written in the limit $t\gg t_{\rm dec}$. 
Compared to the forward shock case, it can be seen that 
$F_{\rm max}\propto t^{-1}$ 
and that $\nu_m$ decays less rapidly (as $t^{-1}$ instead of $t^{-3/2}$).
For a wind environment, the cooling frequency has the same power law
dependence, $\nu_c\propto t^{1/2}$. 
From these expressions the flux can be computed for the different
possible radiative regimes
(Sari, Piran \& Narayan, 1998), the results being given in Appendix A. 

Let us for example take the following values of the parameters: 
$\Gamma_2=1$, $\epsilon_e=\epsilon_B=0.33$, 
$\zeta=10^{-2}$, $p=2.5$, $A_*=0.5$ and 
$t_{\rm dec}=100$ s. Then, the transition from fast to slow cooling occurs
at $t=1.3(1+z)$ day. Now adopting 1 keV and 2 eV 
as typical energies for
the X-ray and visible bands (i.e. $\nu_X=2.4\,10^{17}$ Hz
and $\nu_V=4.8\,10^{14}$ Hz) and a redshift $z=2$, it appears that 
after only a few seconds $\nu_X$ becomes larger than $\nu_m$ and 
then remains larger than $\nu_c$ in the slow cooling regime. 
The corresponding temporal slope is $\alpha_X=(2p+1)/4=1.5$.
At the visible frequency, we initially have $\nu_c<\nu_V<\nu_m$ and therefore
$\alpha_V=0.75$. The visible frequency crosses $\nu_m$ at $t=2.6$ h
(in the fast cooling regime) and then $\nu_c$ (in slow cooling) at very
late times. A break from $\alpha_V=0.75$ to 1.5 is expected at 
$t=2.6$ h.

Since these predicted slopes are only valid in the asymptotic 
regime where $t\gg t_{\rm dec}$ we have performed a numerical 
simulation with the burst parameters given above except for the fraction 
$\zeta$ of accelerated electrons which is varied between 0.003 and 0.03.
We assume in
addition that $E_{53}=1$ and adopt
a rest frame reddening $A_V=0.5$ in the burst host galaxy. 
The resulting X-ray and visible light curves 
are shown in Fig.4. For $\zeta=3\,10^{-3}$ and $10^{-2}$ they exhibit chromatic breaks.
The break in X-rays is a consequence of the dynamics of the reverse 
shock (it is already present in the bolometric light curve) while the
break in the visible is a spectral break (when $\nu_V$ crosses $\nu_m$).
The cases with $\zeta=3\,10^{-3}$, $10^{-2}$ and $3\,10^{-2}$ are very
similar to the early afterglow light curves of respectively
GRB 050802, GRB 050922c and GRB 050801 (see Panaitescu et al, 2006
and Panaitescu, 2006).

The subsequent evolution of the afterglow 
will depend on the behavior of $\epsilon_e$ and $\epsilon_B$ in the
forward shock. If they increase enough with time the forward shock
contribution will eventually dominate but the moment of the transition is difficult
to estimate in the absence of any reliable physical model for the possible
variations of the shock microphysics parameters. If the forward shock takes
over after about one day, the multiwavelength fits of GRB afterglows obtained
in the pre-SWIFT era will remain valid but the early afterglow 
will be explained by the reverse shock. At the transition,
a change of slope or the presence of a bump may however be expected.
While such accidents have been observed in some bursts they do not seem to be
a generic feature of GRB afterglows.

A much more radical point of view can still be adopted: it would to 
suppose that in some cases the forward shock never takes over so that the afterglow
is entirely produced by the reverse shock! The results shown in Fig.4
seem to indicate that this possibility should not be excluded a priori 
even if, clearly, considerable work will be needed to confirm it. 
As for the forward shock hypothesis it will have to be confronted to a large
amount of multiwavelength afterglow data and show that it can 
provide a consistent picture for their interpretation.
\begin{figure*}{}
\begin{center}
\begin{tabular}{cc}
\resizebox{0.33\hsize}{!}{\includegraphics{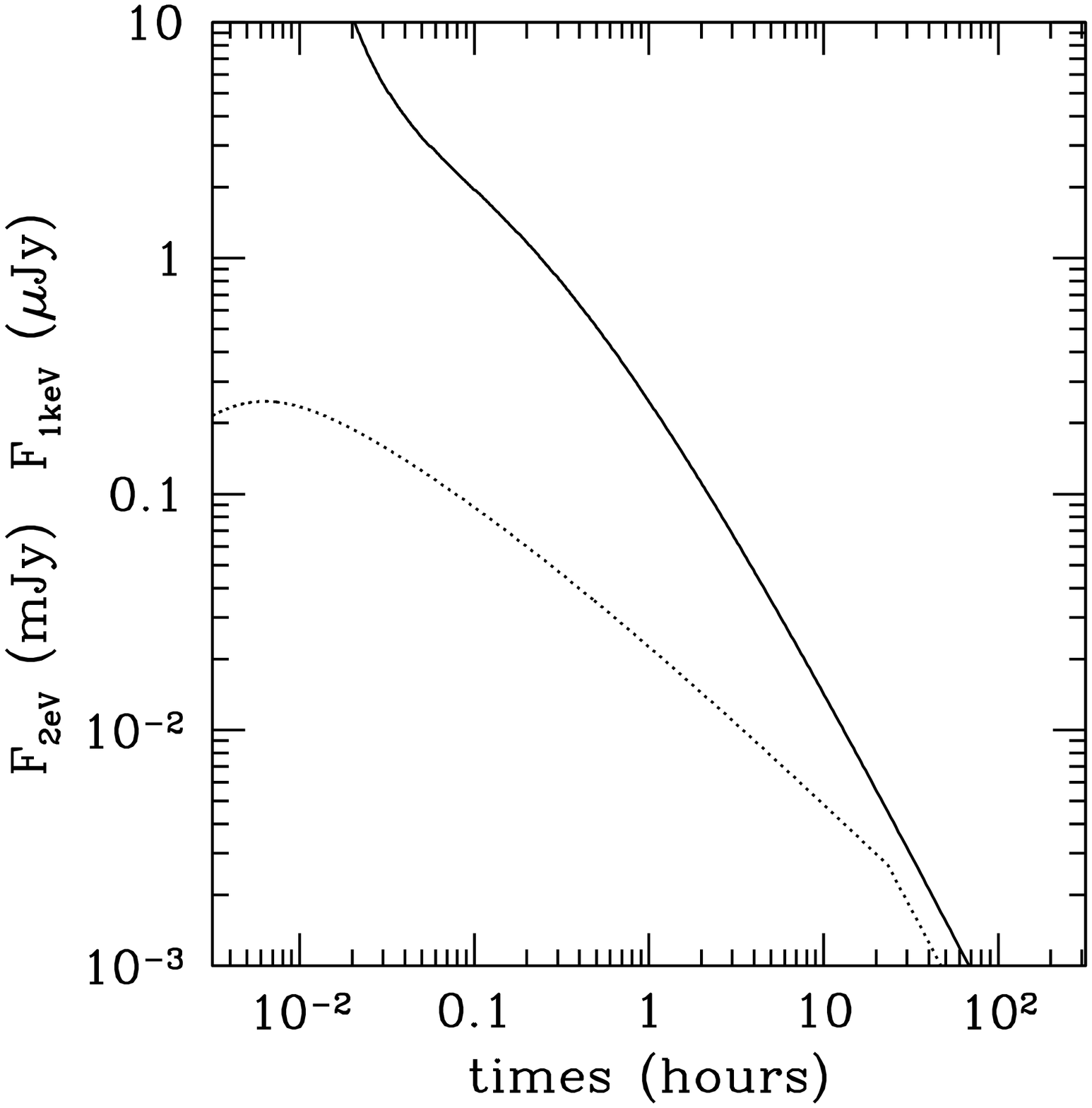}}&
\resizebox{0.33\hsize}{!}{\includegraphics{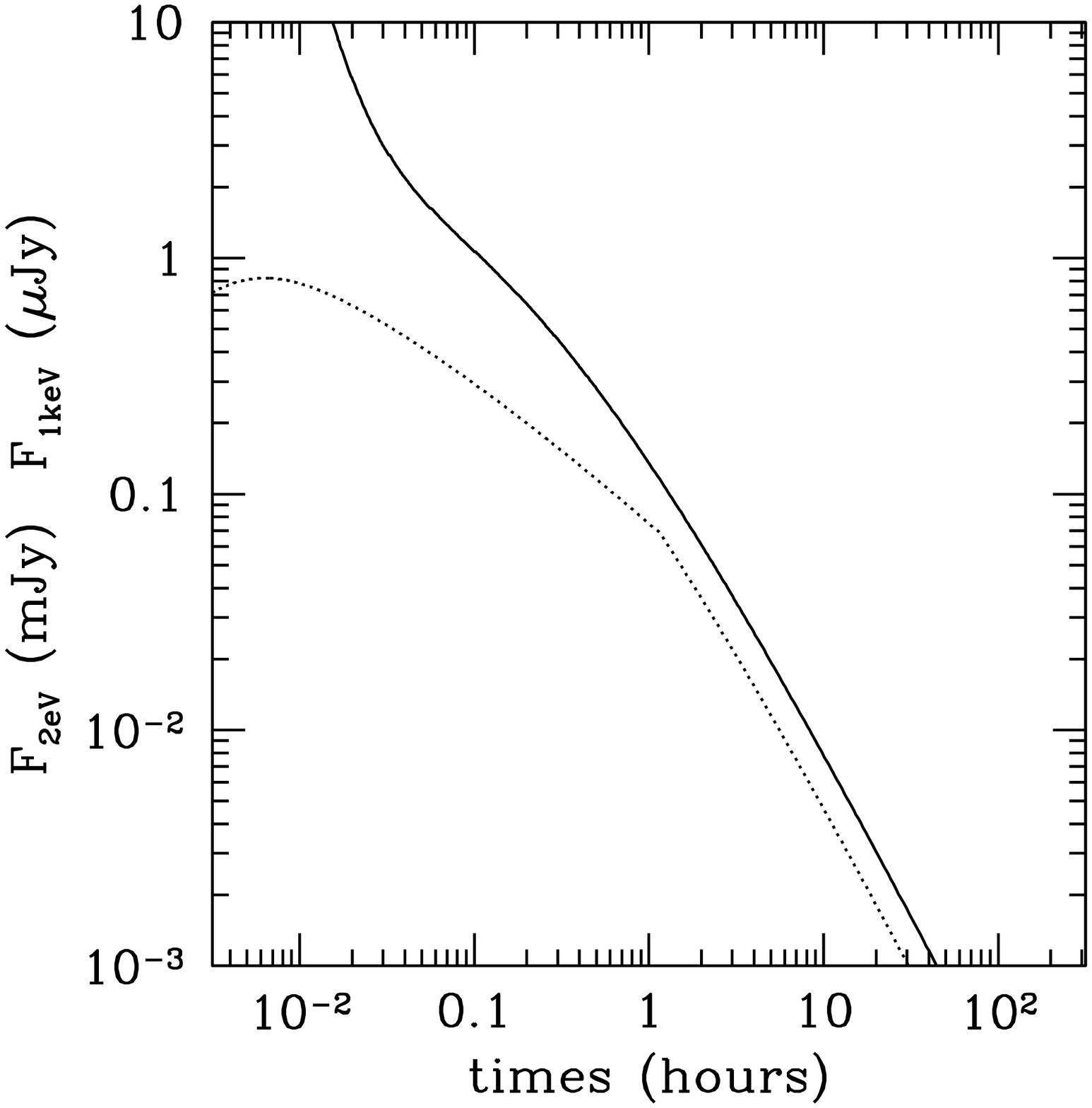}}
\resizebox{0.33\hsize}{!}{\includegraphics{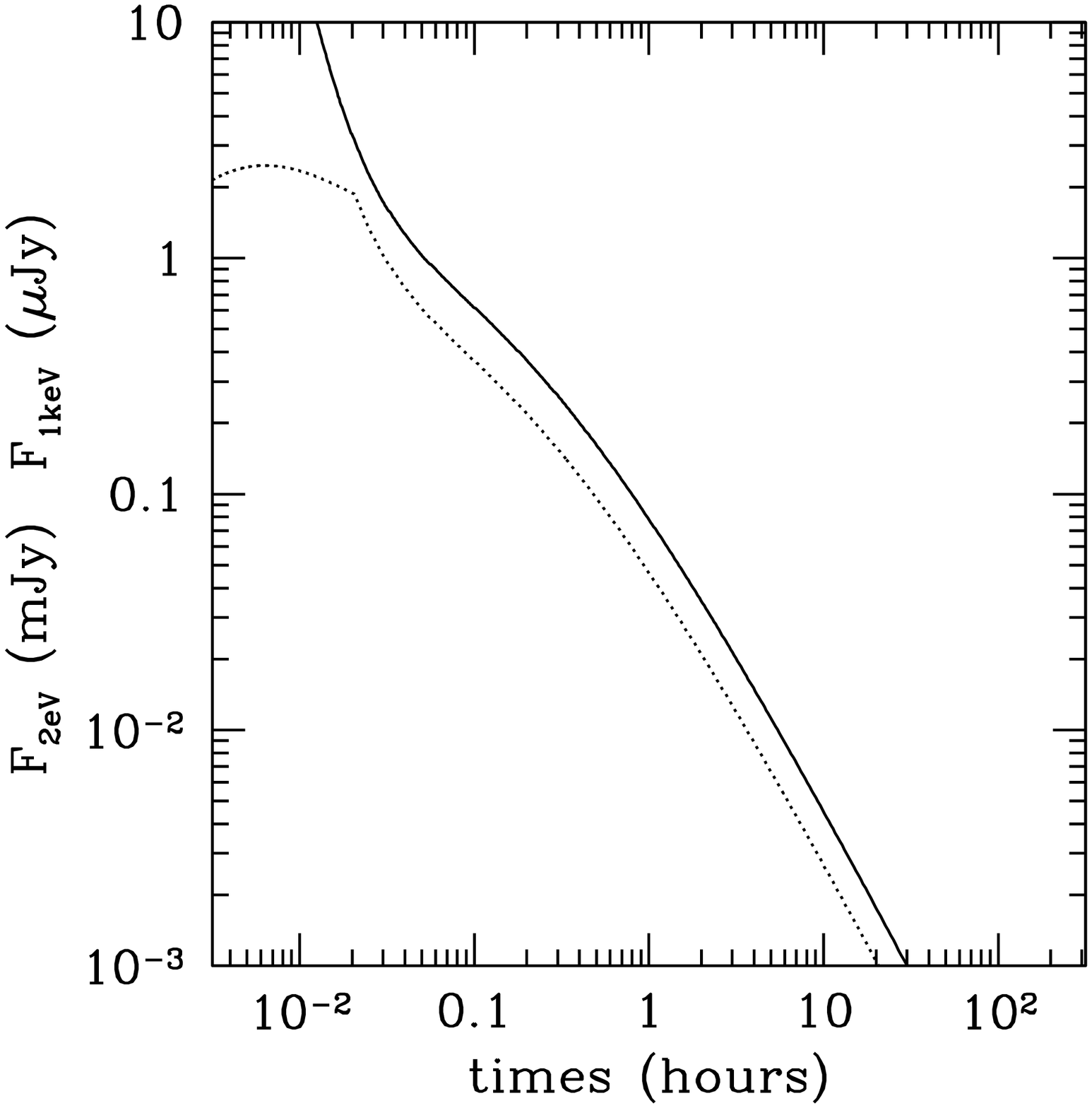}}\\
\end{tabular}
\end{center}
\caption{Early afterglow light curves produced by the reverse shock for
$\epsilon_e=\epsilon_B=0.33$, $p=2.5$, $\Gamma_2=1$, $A_*=0.5$, 
$t_{\rm dec}=100$ s, $A_V=0.5$ and, from left to right, $\zeta=3\,10^{-3}$,
$10^{-2}$ and $3\,10^{-2}$ (see text for details). The full (resp. dotted) line is the
X-ray (resp. the visible) afterglow. Compare these results to the
early afterglow light curves of respectively 
GRB 050802, GRB 050922c and GRB 050801 
as shown in Panaitescu (2006).}
\end{figure*}
\section{Conclusion}
We have developed a simplified model which enabled us to follow simultaneously the 
dynamics of the internal, external and reverse shocks in GRBs. We were mainly interested
by dissipation in the reverse shock when the Lorentz factor in the material which 
is ejected at late times by the source decreases to small values, $\Gamma_{\rm f} < 10$.
The propagation of the reverse shock then extends over quite a long time needed to
decelerate the fast moving part of the ejecta down to 
$\Gamma\sim \Gamma_{\rm f}$. We have obtained 
the dissipated power as a function of observed time for different 
burst environments
(wind or constant density).
Its evolution shows a stri\-king resemblance 
with the early afterglow light curves observed by SWIFT,
especially in the 
wind case. 
However the reverse shock contribution is normally expected 
at low energy and to appear 
in X-rays it requires a transfer of the dissipated power 
to only a small fraction 
($\zeta\lsim 10^{-2}$) of the electron  population. If this is possible, SWIFT XRT
observations could be better explained by the reverse shock than by the standard afterglow
produced by the forward shock. 

We have also computed the optical emission from the reverse 
shock. The comparison with the X-ray light curve often reveals the
presence of chromatic breaks during the first hours. Such breaks 
have been observed and are 
difficult to explain with the standard model where the afterglow
comes from the forward shock.
We have finally even proposed that in some cases the entire afterglow
could be produced by the reverse shock. We fully understand that, 
to be validated, this non standard view still has to show that it
can successfully explain multiwavelength observations of a reasonable
sample of GRB afterglows. We aim to perform these necessary tests in a work
in preparation.

\appendix
\section{}
Using Eq.(37) for $F_{\rm max}$, $\nu_m$ and $\nu_c$ 
we give the expressions for the flux at a given frequency in  
the fast and slow cooling regimes:
\vskip 0.3cm\noindent
{\it Fast cooling}\vskip 0.2cm\noindent
1) $\nu<\nu_c$ 
\begin{eqnarray}
F_{\nu}& = &F_{\rm max}\left({\nu\over \nu_c}\right)^{1/3} \nonumber \\
& = &10^{11}\,{(1+z)^{4/3}\over D_{28}^2}\,{(\zeta E_{53})(\epsilon_B A_*)\over 
t_{\rm dec}^{1/6}\Gamma_2^{5/3}}\,\nu_{17.4}^{1/3}\times t^{-7/6}\ \ {\rm mJ}
\end{eqnarray}
2) $\nu_c<\nu<\nu_m$ 
\begin{eqnarray}
F_{\nu}& = &F_{\rm max}\left({\nu\over \nu_c}\right)^{-1/2} \nonumber \\
& = &7.7\,10^{3}\,{(1+z)^{1/2}\over D_{28}^2}\,{(\zeta E_{53})\,t_{\rm dec}^{1/4}\over 
(\epsilon_B A_*)^{1/4}}\,\nu_{17.4}^{-1/2}\times t^{-3/4}\ \ {\rm mJ}
\end{eqnarray}
3) $\nu>\nu_m$ 
\begin{eqnarray}
F_{\nu}& = &F_{\rm max}\left({\nu_m\over \nu_c}\right)^{-1/2} 
\left({\nu\over \nu_m}\right)^{-p/2}\nonumber \\
& = &1.3\,10^{4}\times 0.36^{p/2}\,{(1+z)^{1-p/2}\over D_{28}^2}\,
(\zeta E_{53})\,(\epsilon_B A_*)^{{p-2\over 4}}\,t_{\rm dec}^{1/4}\nonumber\\
&\times&\left({\epsilon_e\over \zeta}\right)^{p-1}\,\left({p-2\over p-1}\right)^{p-1} 
\nu_{17.4}^{-p/2}\times t^{-{2p+1\over 4}}\ \ {\rm mJ}
\end{eqnarray} 
\vskip 0.3cm\noindent
{\it Slow cooling}\vskip 0.2cm\noindent
1) $\nu<\nu_m$ 
\begin{eqnarray}
F_{\nu}& = &F_{\rm max}\left({\nu\over \nu_m}\right)^{1/3} \nonumber \\
& = &2\,10^{8}\,{(1+z)^{4/3}\over D_{28}^2}\,{(\zeta E_{53})(\epsilon_B A_*)^{1/3}\over 
\Gamma_2}\nonumber\\
&\times& \left({\epsilon_e\over \zeta}\right)^{-2/3}\,\left({p-2\over p-1}\right)^{-2/3} 
\nu_{17.4}^{1/3}\times t^{-2/3}\ \ {\rm mJ}
\end{eqnarray} 
2) $\nu_m<\nu<\nu_c$
\begin{eqnarray}
F_{\nu}& = &F_{\rm max}\left({\nu\over \nu_m}\right)^{{1-p\over 2}}
\nonumber \\
& = &1.4\,10^{8}\times 0.36^{p-1\over 2}\,{(1+z)^{3-p\over 2}\over D_{28}^2}\,
{(\zeta E_{53})\,(\epsilon_B A_*)^{{p+1\over 4}}\over \Gamma_2}\nonumber\\
&\times& \left({\epsilon_e\over \zeta}\right)^{p-1}\,\left({p-2\over p-1}\right)^{p-1} 
\nu_{17.4}^{(1-p)/2}\times t^{-{p+1\over 2}}\ \ {\rm mJ}
\end{eqnarray}
3) $\nu>\nu_c$
\begin{eqnarray}
F_{\nu}& = &F_{\rm max}\left({\nu_m\over \nu_c}\right)^{{p-1\over 2}}
\left({\nu\over \nu_c}\right)^{-{p\over 2}}
\nonumber \\
& = &1.3\,10^{4}\times 0.36^{p/2}\,{(1+z)^{1-p/2}\over D_{28}^2}\,
(\zeta E_{53})\,(\epsilon_B A_*)^{{p-2\over 4}}\,t_{\rm dec}^{1/4}\nonumber\\
&\times& \left({\epsilon_e\over \zeta}\right)^{p-1}\,\left({p-2\over p-1}\right)^{p-1} 
\nu_{17.4}^{-p/2}\times t^{-{2p+1\over 4}}\ \ {\rm mJ}
\end{eqnarray}
In all these expressions the frequency (in observer frame) is in unit of $10^{17.4}$ Hz, 
corresponding  to 1 keV.

\section*{Acknowledgments}
The authors would like to thank John Eldridge for a careful reading of the manuscript.

\end{document}